# Integrated Hybrid Optical Networking (IHON) /Fusion for 5G Access Networks

Dawit Hadush Hailu*[1], Gebrehiwet Gebrekrstos Lema[2] and Gebremichael T. Tesfamariam[3]

[1,2,3] School of Electrical and Computer Engineering, Ethiopian Institute of Techonology-Mekelle (EiT-M), Mekelle University, Mekelle, Tigray, Ethiopia
*Corresponding author, e-mail: dawithadush@gmail.com

**Abstract:**
Today, deployment of optical fiber has offered large transmission capacity which cannot be efficiently utilized by the electronic switches. Rather, Integrated Hybrid Optical Network (IHON) is a promising approach which combines both packet and circuit switching techniques. As a result, it achieves efficient utilization of the bulk capacity and guarantees absolute Quality of Service (QoS) by optimizing the advantages of the two switching schemes while diminishing their disadvantages. Transpacket has developed a Fusion node implementing IHON principles in Ethernet for the data plane. Hence, this paper investigates and evaluates IHON network for 5G access networks. The simulated results and numerical analysis confirm that the Packet Delay Variation (PDV), Delay and Packet Loss Ratio (PLR) of Guaranteed Service Transport. (GST) traffic in IHON network met the requirements of 5G mobile fronthaul using CPRI. The number of nodes in the network limits the maximum separation distance between Base Band Unit. (BBU) and Remote Radio Head (RRH), link length; for increasing the number of nodes, the link length decreases. In addition to this, we verified how the leftover capacity of fusion node can be used to carry the low priority packets and how the GST traffic can have deterministic characteristics on a single wavelength by delaying it with Fixed Delay Line (FDL). For example, for $L^{SM}_{1GE}=0.3$ the added Statistical Multiplexing (SM) traffic increases the 10GE wavelength utilization up to 89% without any losses and with SM PLR=$1E^{-03}$ up to 92% utilization.

## 1. Introduction

In recent years, the dynamic growth of fiber deployment has brought more transmission capacity in to hand, which is beyond the processing capacity of electronic switches [1]. As a result, fiber switching technologies [2], have been introduced to overcome the electronics processing and switching bottlenecks. Even though optics offers bulk capacity, the utilization is inefficient due to slow adaptation of wavelength switching for burst traffic. All-optical switching is believed to enhance the efficiency, but it is not practically deployed [1]. Hence, a new adaptive design approach, the hybrid optical architecture [1], was proposed: it combines the best features of both optical circuit and packet switching while diminishing their demerits to provide better performance and cost reduction [1]. Integrated Hybrid Optical Network (IHON) is one type of hybrid network which completely integrates both circuit and packet technologies, and uses the same wavelength to transmit both traffic. As a result, IHON offers guaranteed Quality of Service (QoS) and efficient utilization of the transmission capacity through effective management.

Managing different technologies separately increases both Operational Expenses (OPEXs) and Capital Expenses (CAPEXs) [3]. To overcome such challenges, Software-Defined Network (SDN) [4], was adopted: SDN separates the data and control planes, thus the network functions and protocols are made programmable. The main principle in SDN, is abstraction of the data plane using controller, and present it to applications as a virtual entity. As a result, the underlying infrastructure merely forwards traffic based on rules from the central controller. Moreover, the SDN scenario controls network resources and data transmission across heterogeneous domains, thus it avoids interoperability problems between Network Elements (NEs) from different vendors [4]. To efficiently utilize the growing transmission capacity, Transpacket has developed Fusion solution (i.e., H1 nodes) [5], based on the principles of IHON. H1 offers efficient capacity utilization while providing absolute QoS for critical traffic. So far, NETCONF-based Network Management System (NMS) [6], is implemented for fusion networks to supervise and manage different traffic flows.

With packet based network realization, the performance of H1 node/Fusion node issue has been one of the biggest challenges. As a result, several continuous researches have been made towards an IHON network for future networks. Using Ethernet in the fronthaul [7] has been proposed to take some advantages: lower cost equipment, shared use of lower-cost infrastructure with fixed access networks, obtaining statistical multiplexing, and optimized performance. Despite of their attractive advantages, Ethernet also comes with their own challenges: achieving low latency and jitter to meet delay requirements, and ultra-high bit rate requirements for transporting radio streams for multiple antennas in increased bandwidth [7]. For the above reasons, the current fronthaul/access networks are increasingly integrating more cost-effective packet switched technology, especially Ethernet/Internet technologies. In addition to standard Ethernet switch used by Ethernet, there is another node, H1, developed by transpacket [8] employed in IHON fusion solution. Fusion solution/IHON that uses standard Ethernet technology rather than all-optical switching technology provides the fusion properties of circuit and packet switching network in packet network. It enables Ethernet transport and ensures strict QoS for Guaranteed Service Transport (GST) traffic, and optimize resource utilization by introducing Statistical Multiplexing (SM) traffic in the unused capacity.

A number of researches have done in optical packet switched networks [9-15]. The papers focused on optimizing PLR by using different Quality of Service (QoS) differentiation policies and scheme of transport networks. The main focus of this paper is to analyzing the different QoS performance metrics that are vital in evaluating the IHON node which will be employed in Ethernet-based access transport network. To the best of author's knowledge, the IHON node performance is evaluated using simulation for the first time.

In this study, we are inspired to evaluate IHON/fusion networks by replacing the existing access network with C-RAN using simulation for 5G access networks. The authors have also highlighted the mobile fronthaul network options for 5G network discussed in [16]. In particular we focus on three performance metrics to evaluate the IHON: PLR, Latency and PDV.

## 2. Reasearch Method

The study has collected secondary data from previously published studies. These materials are background research, conference papers, white papers, International Telecommunication Union - Telecommunication (ITU-T) and Internet Engineering Task Force (IETF) standard recommendations. This study has focused on collecting the QoS requirements of IHON LAN.

Generally, Figure 1, the flow used to achieve the objective of the work includes:
**System modeling:** includes the designing, implementing, evaluating and analyzing of a single IHON node for 5G access fronthaul network.
**Performance analysis:** includes performance analysis of IHON node with respect to delay, packet loss ratio and packet delay variation.
**Comparsion:** comparing of the results obtained from the simulator with IEEE 802.1CM standard and IEEE 1904.3 Task Force.
**Interpretation of the results:** the performance analysis of IHON network for 5G access networks, comparison and the interpretation of the results will be explained.

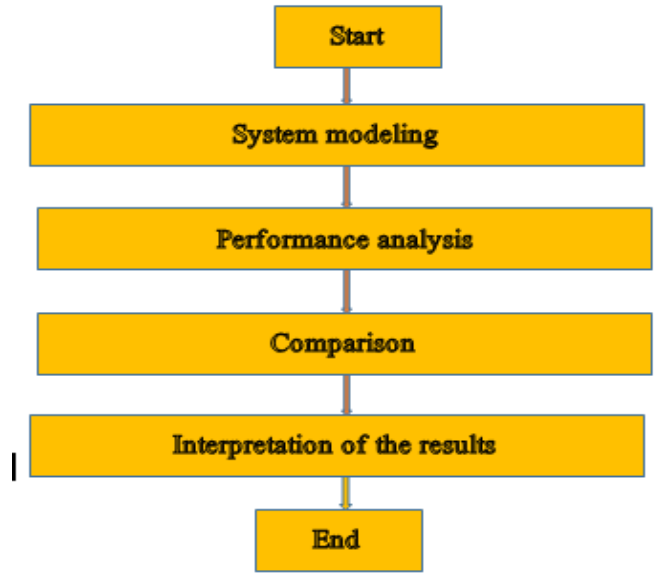

Figure 1: Methodology flow chart

## 3. System Model for Simulation

To study the performance of IHON node while transporting GST and SM traffic, the traffic has been transported and analyzed on a 10GE output wavelength. Figure 2 illustrates how these traffic has been generated, and utilize the output link. Firstly, the GST and SM traffic were generated. After generating, the GST traffic was sent to the 10 GE port of the IHON node while the SM traffic was queued in a buffer until a suitable gap was detected. Secondly, both the GST and SM traffic was processed and sent out through the same output port.

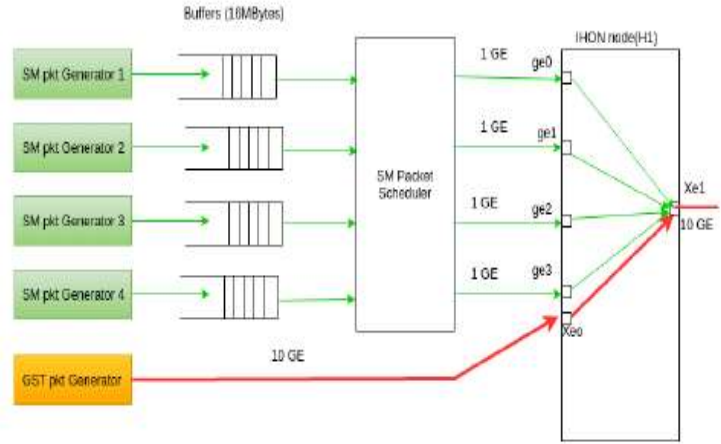

Figure 2: Diagram illustrating how the IHON node is connected to Packet generators for measuring the performance metrics.

### 3.1. Simulation Model for IHON Node

In this work, we simulate IHON node based on non-preemptive scheduling using the programming language called simula/Demos programming. The software is designed and implemented as a full scale general purpose programming. As its name implies, it is an object-oriented simulation language developed at the Norwegian Computing Center in Norway for designing of simulation entities. It has also been used in simulating Very Large-Scale Integration (VLSI) designs, process modeling, and other applications [17].

#### 3.1.1. Traffic patterns

In the IHON node implementation, the arrival processes of GST and SM traffic are described by negative exponential distributions, where as their packet lengths are described by deterministic and uniform distributions respectively. The GST packets are allowed to travel into a network following reserved wavelength available so that they do not experience any packet loss. The rate of GST traffic source is defined using negative exponential distribution with a mean value which is a function of capacity of the output port and the length of GST packet. Hence, it ensures the wanted load from GST traffic on single wavelength. Unlike GST packets, since SM traffic has no pre-established path it may experience a packet loss and is inserted in suitable gaps between GST packets. And its mean value is a function of the channel capacity and length of SM packets drawn from the distribution.

Similarly, negative exponential distributions are used to describe the arrival processes for HP and LP traffic in Ethernet switch implementation. The packet length distribution used to describe the length of HP and LP packet is deterministic distribution. HP traffic is given highest priority, and LP traffic is given low priority. The mean value for HP and LP traffic in Ethernet switch follow the same pattern as GST and SM traffic in IHON node.

### 4. Simulation Parameters

Table I and Table II present the set of parameters which have been used during the simulation and the list of notations used for different traffic loads on different interfaces respectively.

TABLE I: Simulation parameters used in the analysis of performance metrics of LP and HP packets.

| Parameters | Value |
| --- | --- |
| Seed values | 907 234 326 104 711 523 883 113 417 656 |
| Output link capacity | 10 Gb/sec |
| Minimum SM length | 40 Bytes |
| Length of GST packet | 1200 Bytes |
| Maximum SM length | 1500 Bytes |
| Load of GST traffic | Varies |

| Load of SM traffic | Varies |
|---|---|
| Buffer size | 16 MByte |
| Number of packets | 40,000 |

TABLE II: Notation of parameters used in the simulation result analysis.

| Description | Notations |
|---|---|
| The load of SM traffic on 1 Gb/s interface | $L_{1GE}^{SM}$ |
| The load of GST traffic on 10Gb/s interface | $L_{10GE}^{GST}$ |
| The load of GST and SM traffic on 10 Gb/s interface | $L_{10GE}^{T}$ |

## 5. Results and Analysis

In this work, we considered PDV, PLR, and average latency as a performance metrics to evaluate performance of IHON node.

### 5.1. IHON Performance

#### 5.1.1 GST Traffic Performance

In this part, the performance of GST traffic is presented with respect to: average latency, PLR, and PDV.

**Average Latency**

In IHON node, latency refers to the delay when the first bit of the packet gets into the IHON node until the first bit of the packet leaves the FDL. This parameter greatly affects how usable the nodes as well as communication are.

Every sub-simulation returns an average value for latency of GST packet, obtained by averaging all the delays experienced by every single GST packet during the sub simulation. By varying the system load ($n*L_{1GE}^{SM} + L_{10GE}^{GST}$), we observed the average latency of GST traffic. The simulated average latency of GST traffic was constant with a value of 1.2 $\mu$sec, which equals to the FDL time. It shows that the 10GE GST traffic was shown to be independent of the added SM traffic and its load. This is due to the fact that the GST traffic is transmitted with absolute priority. Note that the FDL time was set to the service time of maximum packet length of SM packet, $\delta$ =1.2 $\mu$sec.

The average latency of GST traffic isn't affected by the system load and insertion of SM traffic. Thus, the result reveals that its average latency is constant regardless of the node congestion.

**Packet Delay Variation**

As per ITU Y.1540, "PDV is the variation in packet delay with respect to some reference metrics (minimum delay in this work)". The service quality and PDV tolerability of an application are highly influenced by PDV.

By recording the time when the first bit of GST packets arrived at the delay line, and when the last bit of the packet left the start of the FDL, the maximum and minimum delay of GST packet is computed. Using this concept, the result of the simulator showed that the packet delay variation of GST packets was zero. This is because the traffic aggregation of SM traffic on the top a 10 Gb/s wavelength is done without introducing PDV to the GST traffic of following the wavelength. Furthermore, the inter-packet gap between the GST streams is preserved, and the GST streams are sent out precisely as it arrived. Consequently, it results in zero packet delay variation. Like the average latency, the PDV of the GST traffic isn't affected by the system load and insertion of SM traffic.

And also, GST packets in IHON node undergo a fixed delay of 1.2 $\mu$sec, corresponding to the service time of a maximum length SM packet. At the output wavelength, all GST packets experience this fixed delay. Hence, it doesn't introduce PDV to GST traffic.

**Packet Loss Ratio**

When one or more transmitted packets fail to arrive at their destination, packet loss occurs. In IHON node, it is typically caused by blocking and congestion. Since different applications have different PLR tolerability, it has noticeable effects in all types of communications. Packet loss is measured as a percentage of the number of lost packets with respect to the total transmitted packets.

The simulation result showed that the PLR of GST traffic was zero, i.e. all GST traffic generated by the source were received at the output port of the IHON node. This is because GST packets pass through FDL, which gives time to the monitoring module to calculate the gap length between GST packets. SM packets are scheduled only if the gap is sufficient to transmit the packet. Hence, IHON nodes avoid losing of GST packets.

### 5.1.2. SM Traffic Performance

By varying the load of GST traffic, we observed the performance of SM traffic for a fixed SM load.

**Average Latency**

For low SM load, Figure 3a, the average latency increases from 1.2 $\mu$sec to 18 $\mu$sec when the GST load, $L^{GST}_{10GE}$ was increased from 0.1 to 0.89 with an increasing interval of 0.1. However, when we increase $L^{SM}_{1GE}$ from 0.1 to 0.3, Figure 3b, the average latency was further increased. Increasing more SM traffic, after the system load 0.8, will cause a buffer overflow; this leads the average latency to increase exponentially and causes a packet loss as shown in Figure 5.

As can be seen from Figure 3, the average latency of SM traffic is increased for increasing the GST load. The increment of GST load adds more traffic to the output wavelength which in turn increases the waiting time of SM packet at the node. This means that increasing GST load will decrease the chance of SM traffic to be inserted. Generally, the higher the GST traffic load, the longer the average latency of SM traffic.

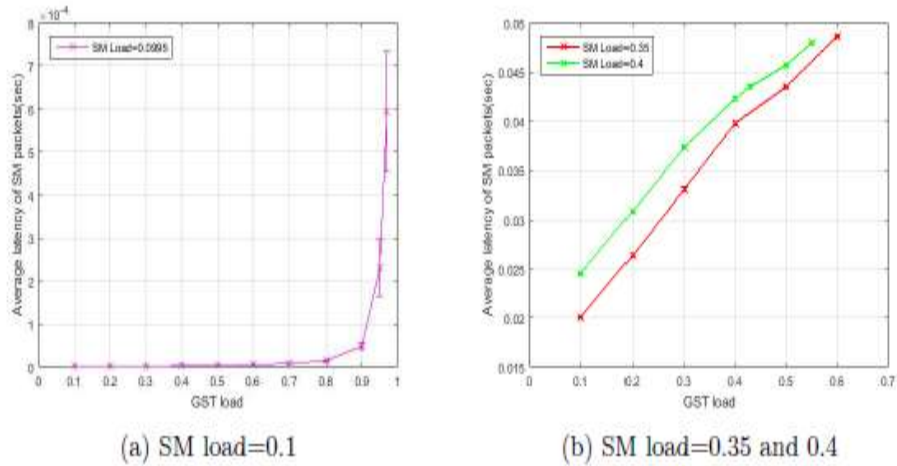

Figure 3: Average latency of SM traffic as function of GST load for SM load=0.1, 0.35 and 0.4.

When the system load is low, then the buffers inside the node will never be full. In this simulation, since we are using variable SM packet length (40 to 1500 bytes), the average latency depends on the packet length. It means that the latency of sending a 40-byte packet and a 1500 byte is different depending on the packet length.

**Packet Delay Variation**

As it has been mentioned, PDV influents service quality of an application. Thus, the average PDV of SM packets were acquired from the simulation for analysis. The resulting Figure 4 shows that when the GST load of the system load traffic increased, the PDV has increased from 20.1 $\mu$sec to 411.232 $\mu$sec. This indicates that the PDV of SM packet is influenced because of the service classification, i.e. traffic load and SM traffic scheduling algorithm [18].

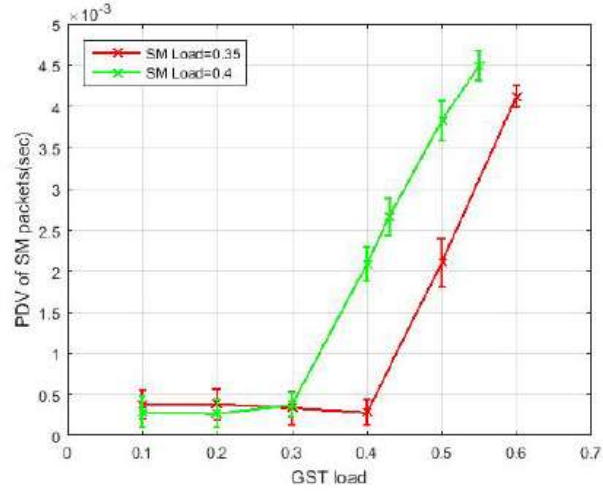

Figure 4: Packet delay variation of SM traffic as function of GST load.

**Packet Loss Ratio**

One of the most important factors in the analysis of IHON node performance for SM traffic is illustrated in Figure 5. Based on the figure, no PLR was observed within system load interval [0, 0.89]. When the system load reaches 0.89, SM packets start getting dropped. From that point onward buffer overflow causes the loss to increase exponentially.

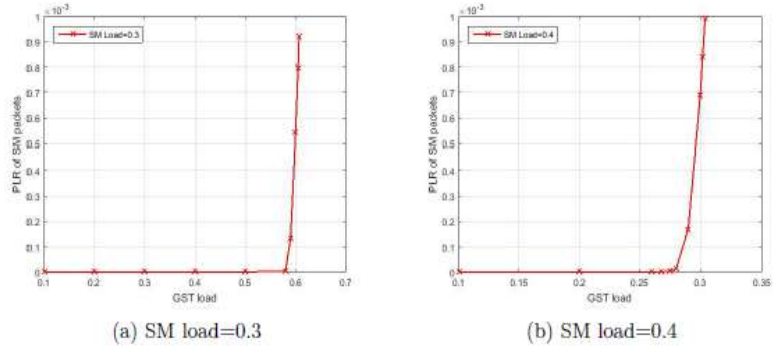

Figure 5: PLR of SM traffic as function of GST load.

For $L^{SM}_{1GE}$=0.3, the added SM traffic increases the 10GE wavelength utilization up to 89% without any losses and with SM PLR=$1E^{-03}$ up to 92% utilization.

### 5.2. *Evaluation of IHON Network for Access Networks*

**Average Latency**

The first performance metric used for evaluation of IHON network for access network C-RAN is average latency. According to IEEE 802.1CM and Table III, the maximum E2E latency between BBU and RRH required for mobile fronthaul is specified as 50 $\mu$sec (including fiber length and PDV), and the latency budget is 5$\mu$sec when cable propagation is excluded. Within this 50 $\mu$sec latency, the maximum separation distance between BBU and RRH (link length) is limited and depends on the number of nodes in the network. To calculate the link length as function of number of nodes (N), the following equation has been used:

Maximum E2E latency ($L_{total}$):
$$L_{total} = N \times (D_{node} + Delay\_PDV_{node}) + D_T \times Link_{length} \qquad (1)$$

where, N = Number of nodes.
$D_{node}$= Latency in a single node, 1.2$\mu$sec (obtained from the simulation).
*Delay_P DV*$_{node}$=PDV in a single node, 0$\mu$sec (obtained from the simulation).
$D_T$ =Transmission latency, 5$\mu$sec/km.

$L_{total}$= Maximum E2E latency, 50$\mu$sec.
$Link_{length}$= Maximum link length.

Table III: Maximum link length and number of nodes in IHON network to meet the fronthaul requirements for GST traffic where $L_{total}$=50$\mu$sec, $D_T$ =5$\mu$sec/km, and $D_{node}$=1.2$\mu$sec.

| N | Total $D_{node}$ ($\mu$sec) | $Link_{length}$(Km) |
|---|---|---|
| 2 | 2.4 | 9.52 |
| 3 | 3.6 | 9.28 |
| 4 | 4.8 | 9.04 |
| 5 | 6 | 8.8 |
| 6 | 7.2 | 8.56 |

Table III presents the relationship between the number of nodes in IHON fronthaul network and the maximum separation distance between BBU and RRH for GST traffic. In the table, the second column indicates the overall delay in the nodes, whereas the third column is the link length between BBU and RRH in the network. The table shows that for increasing the number of nodes in the network, the link length decreases. For instance, for IHON network with 3 nodes require 9.28 Km fiber link length. Similarly, for IHON network with 4 nodes require 9.04 Km fiber link length.

The simulation results and Table III proved that IHON networks are capable of carrying radio signals over packet-based fronthaul network provided that the number of nodes in the table corresponds to its respective link length. Unfortunately, obtaining the average latency of BE or SM traffic that meets fronthaul requirement is not so straightforward, since this traffic depends on a load of GST traffic.

**Packet Delay Variation**

Another performance used for evaluation of IHON network for a fronthaul network is PDV. The maximum PDV specified is 5 $\mu$sec or 10% of E2E latency. The simulated result of PDV for GST traffic was zero. Thus, the GST traffic class meets the fronthaul requirement in PDV comparison. The fronthaul requirement is higher than peak PDV of GST traffic for any system load in the interval [0, 0.99]. Consequently, of the three performance metrics (average latency, PDV, PLR), the IHON network performs best in PDV for GST traffic. It has no restriction in wavelength utilization of IHON network.

**Packet Loss Ratio**

At last, the performance evaluated is PLR. Fronthaul networks have a very strict PLR requirement which is in the interval [$10^{-6}$, $10^{-9}$]. As described, GST packets experience no packet loss regardless of the network congestion. So, the GST traffic is transported through the IHON fronthaul network with absolute priority. In our simulation, GST packet loss wasn't observed at any system load, while SM packets getting dropped for system load beyond $L_{10GE}^T$=0.89 for SM load=0.3. With system load in the interval [0, 0.99], GST traffic class meets the fronthaul requirement, and can be served properly in IHON network.

## 6. Conclusion

In this work, the use of IHON networking has been evaluated and investigated to employ it in 5G access networks. The simulation and numerical analysis result verified the effectiveness of IHON network. In the analysis of IHON node, the performance of delivering high quality of service was approved and reflected on GST and fits the technical specification of fusion/IHON node. On the other hand, SM traffic is suitable for transporting time insensitive information. The recommendation from the IEEE 802.1CM standard and other articles were considered for examining and investigating how IHON access network networks should provide the required quality of service.

Other important points that this work has measured and evaluated:
1. The important design concepts of fusion node have been confirmed.
2. The fusion node aggregation of multiple 1GE SM traffic and 10GE on a single wavelength has been achieved without affecting the deterministic nature of circuit switching.
3. The simulation result confirms that the average latency of GST packet is independent of the system load and experiences a zero packet delay variation in a node.
4. The efficient utilization of bandwidth has been achieved by inserting suitable SM traffic in the computed gap between GST packets.

**ACKNOWLEDGEMENTS**

The authors would like to acknowledge Ethiopian Institute of Technology- Mekelle, Ethiopia, for supporting this study. Portions of this work were presented and published in thesis form in fulfillment of the requirements for the MSc. degree for one of the author's, Dawit Hadush Hailu from NTNU [19].

**CONFLICT OF INTEREST**

The authors declare that there are no conflicts of interest.

BIBLIOGRAPHY

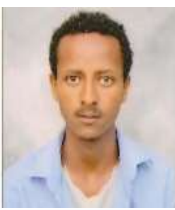

***Dawit Hadush Hailu*** received his BSc degrees in Electrical and Computer Engineering from Ethiopian Institute of Technology-Mekelle (EIT-M), Mekelle University, Mekelle, Ethiopia, in 2013 and his MSc degree from Norwegian University of Science and Technology (NTNU), Trondheim,

Norway, in 2016. He is currently working as Assistant Professor in EiT-M, Mekelle University. His research interest is in the area of networking with special emphasis on Cloud Radio Access Network (C-RAN), mobile fronthaul, radar systems, Signal Processing, Software Defined Network (SDN), optical networking and antenna design.

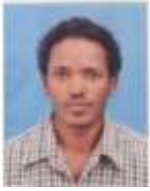

**Gebrehiwet Gebrekrstos Lema** has received his BSc in Electronics and Communication Engineering from Mekelle Institute of Technology (MIT), Mekelle University, in 2010 and his MSc in Communication Engineering from Ethiopian Institute of Technology-Mekelle (EiT-M) in 2015. He was working as a lecturer in EiT-M, Mekelle University and currently he is attending his PhD in TU of Ilmenau, German. His research interest focuses in antenna design, Self-Organized networks, cellular future networks, optimization technique, beam forming, radar systems, mobile and wireless communication, Signal Processing, Data and Computer Networking.

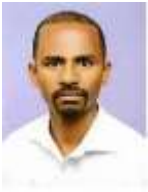

**Gebremichael T. Tesfamariam** (**Dr. –Ing.**) has got his BSc in Electrical Engineering and MSc in Control Engineering from Addis Ababa University in 2001 and 2005 respectively. He studied his PhD in Signal Processing at Technische Universität Darmstadt, Germany in 2013. Currently, he is head of School of Electrical and Computer Engineering, EiT-M, Mekelle University, Ethiopia. His research interests are Signal Processing, Radar Systems, Antenna design and Adaptive control systems.

Table IV: Access network requirements of demanding services and applications based on ITU-T recommendation Y.1541 [20]

| Service Classes | Y.1541 QoS Class | Upper bound PLR | Upper bound Delay | Upper bound jitter | Bandwidth needed |
|---|---|---|---|---|---|
| i. Video streaming | 6 or 7 | $10^{-5}$ | 100 ms or 400 ms | 50 ms | High |
| ii. Video Conversational | 0 or 1 | $10^{-3}$ | 100 ms or 400 ms | 50 ms | High |
| iii. Music streaming | 6 or 7 | $10^{-5}$ | 100 ms or 400 ms | 50 ms | Low to medium |
| iv. Voice conversational | 0 or 1 | $10^{-3}$ | 100 ms or 400 ms | 50 ms | Low |
| v. Interative messaging | 3 (or2) | $10^{-3}$ | 100 ms or 400 ms | Undef. | Low |
| vi. Control traffic | 2 | $10^{-3}$ | 100 ms | Undef. | Low |
| vii. General data transfer | 4 or 5 | $10^{-3}$ or undef. | Undef. | Undef. | Low to High |